# Absorption of Terahertz Radiation in Ge/Si(001) Heterostructures with Quantum Dots

E. S. Zhukova, B. P. Gorshunov, V. A. Yuryev, L. V. Arapkina, K. V. Chizh, V. A. Chapnin,
V. P. Kalinushkin, A. S. Prokhorov, and G. N. Mikhailova

*Prokhorov General Physics Institute, Russian Academy of Sciences, ul. Vavilova 38, Moscow, 119991
Russia email: gorshunov@ran.gpi.ru, vyuryev@kapella.gpi.ru*

The terahertz spectra of the dynamic conductivity and radiation absorption coefficient in germanium–silicon heterostructures with arrays of Ge hut clusters (quantum dots) have been measured for the first time in the frequency range of 0.3–1.2 THz at room temperature. It has been found that the effective dynamic conductivity and effective radiation absorption coefficient in the heterostructure due to the presence of germanium quantum dots in it are much larger than the respective quantities of both the bulk Ge single crystal and Ge/Si(001) without arrays of quantum dots. The possible microscopic mechanisms of the detected increase in the absorption in arrays of quantum dots have been discussed.
**DOI:** 10.1134/S0021364010240033

Artificial low-dimensional nanoobjects—quantum wells, quantum wires, and quantum dots—as well as structures based on them, are promising systems for improvement of existing devices and development of fundamentally new devices of micro and optoelectronics [1]. In addition to the necessity of the development of new technological processes and equipment for manufacturing nanostructures with required parameters, the investigation of the fundamental properties of such structures is also of primary importance. Size quantization of the energy of charge carriers whose spatial motion is limited at scales of about 100 nm or smaller is one of the most striking such properties. In particular, a quantum dot is a zero dimensional object can be considered as an artificial atom with one or more charge carriers (electrons or holes) having a discrete energy spectrum [2]. Arrays of a large number of quantum dots including multilayer heterostructures make it possible to form artificial "solids" whose properties can be controllably changed by varying the characteristics of constituent elements ("atoms") and the environment (semiconductor matrix). Such systems can have a very rich set of physical properties, which are caused by single-particle and collective interactions and depend on the number and mobility of carriers in quantum dots, Coulomb interaction between the carriers inside a quantum dot and in neighboring quantum dots, charge coupling between neighboring quantum dots, polaron and exciton effects, etc. These properties are actively studied during the last two decades. One of the main tools for such investigations is infrared spectroscopy, whose advantages are the contactless character of the measurements and the possibility of the direct observation of transitions between quantized energy states. The specificity of the infrared region is that almost all main interactions in low-dimensional nanostructures (distance between levels, Coulomb interaction between charges in quantum dots, one and multi-particle exciton and polaron effects, plasmon excitations) have the corresponding characteristic energies from several meV to 50–100 meV [3–7].

It is worth noting that almost all performed infrared spectroscopy experiments were devoted to the measurement of the relative characteristics (e.g., relative transmission coefficient) of samples in order to obtain information on the positions of the resonance absorption lines. In this work, we report the first detailed measurements of the *absolute* dynamic conductivity and absorption coefficient in multilayer Ge/Si hetero structures with Ge quantum dots at terahertz and sub-terahertz frequencies ν = 0.03–1.2 THz. Such investigations are important, because it is necessary to know the electrodynamic parameters of nanostructures that can be used in microelectronic systems operating at terahertz frequencies.

The studied structures were multilayer "sandwiches" consisting of a *p*-type Si(100) single-crystal substrate (KDB12 grade) with a resistivity of $\rho = 12$ Ω cm and a thickness of about 500 μm and silicon and germanium deposited layer-by-layer on the surface of the substrate by molecular beam epitaxy in ultrahigh vacuum [8]. The structures were grown in a Riber EVA32 ultrahigh vacuum molecular beam epitaxy chamber integrated with a GPI300 scanning tunneling microscope (STM) [9]. This facility was described in more detail in [8].

When growing the structures with quantum dots, a buffer silicon layer with a thickness of about 100 nm was first deposited on the specially prepared atomically clean surface of a (001) silicon substrate at the temperature $T_{gr} = 530$°C. Then, a germanium layer with the effective thickness $h_{Ge}$ ranging from 4 to 18 Å in various samples was deposited at the temperature $T_{gr} = 350$°C. The conditions of the preparation and structure of the atomically clean Si(001) surface, the parameters of the deposition of Ge and morphological features of the formed arrays of Ge hut clusters [10], and the methods for controlling the epitaxial growth process and the procedure of the STM measurements were described in detail in [8–16]. The deposited Ge layer was coated by a barrier silicon layer with a thickness of 50 nm at $T_{gr} = 530$°C. Five layers of the Ge/Si(001) structure were thus deposited. An additional silicon cap layer with a thickness of 100 nm were finally deposited on the structure at $T_{gr} = 530$°C. The grown layers were undoped. We analyzed several such structures with various $h_{Ge}$ values. The STM images of the surfaces of the germanium layers with various $h_{Ge}$ values are shown in Fig. 1 (the images were processed with the WSxM software [17]), where the evolution of the Ge layer on the Si(001) surface in the process of low-temperature molecular beam epitaxy is seen. Hut clusters on the Ge surface have not been nucleated at $h_{Ge} = 4.4$ Å, and the STM image in Fig. 1a exhibits only the well-known structure of the wetting layer with the $c(4 \times 2)$ or $p(2 \times 2)$ reconstruction inside $M \times N$ blocks [15, 16]. Hut clusters are clearly seen in Fig. 1b for $h_{Ge} = 6$ Å; their density and sizes increase at $h_{Ge} = 8$ Å (see Fig. 1c); clusters with various sizes—completely formed clusters, recently nucleated small clusters, and nuclei with a height of 1 monolayer (ML) over the Ge wetting layer (1 ML ~ 1.4 Å)—are simultaneously seen on the surface [15]. This array is very inhomogeneous both in the sizes of the clusters and in composition; it includes regular pyramidal and elongated wedge-shaped clusters, but wedge-shaped clusters with a large spread in the lengths dominate [8]. The array is most homogeneous at $h_{Ge} = 10$ Å (see Fig. 1d) [19], clusters cover almost the entire surface of the wetting layer, the fraction of small clusters decreases noticeably, and large clusters begin to coalesce. At $h_{Ge} = 14$ Å, most clusters coalesce near their bases (see Figs. 1e, 1f), and the free wetting layer almost disappears from the field of view of the scanning tunneling microscope, but the array consists of individual clusters. At $h_{Ge} = 14$ Å, the coalescence of clusters continues and a transition to the growth of a two-dimensional film of nanocrystalline germanium begins (see Fig. 1g). Finally, at $h_{Ge} = 18$ Å, it is seen that the array of Ge clusters disappears and although the roughness of the surface is still pronounced, the Ge layer grows as a continuous nanocrystalline film (see Fig. 1h). According to Fig. 2, where the density of the clusters in the array is shown as a function of $h_{Ge}$, the density of the wedge-shaped clusters increases when $h_{Ge}$ increases up to 8 Å and, then, decreases slowly, whereas the density of the pyramid shaped clusters decreases exponentially in the process of growth of the array [8]. The total density of the clusters is $3.5 \times 10^{11}$, $5.8 \times 10^{11}$, $5.1 \times 10^{11}$, and $2.3 \times 10^{11}$ cm$^{-2}$ at $h_{Ge} = 6, 8, 10$, and 14 Å, respectively.

When analyzing the terahertz spectra, a structure grown in accordance with the thermal cycle of the growth of the samples with Ge/Si heterostructures was used as a reference sample; Si layers with the corresponding thicknesses were grown at $T_{gr} = 530$°C and Ge was not deposited. Instead of the growth of a Ge film, the sample was cooled to $T_{gr} = 350$°C and was then heated to 530°C with standard rates [8] (the sample cooling time from 530 to 350°C is always much larger than the time taken for the growth of the Ge layer). The sample of a five-layer heterostructure with Ge quantum dots ($h_{Ge} = 10$ Å) grown on high-resistance *p*-Si(100) with $\rho = 1000$ Ω cm (KDB1000) was also studied.

In addition to the samples with Ge quantum dots and the reference sample, we investigated the sample of the Si$_{0.942}$Ge$_{0.058}$/Si(001) heterostructure grown at 680°C (150-nm Si buffer layer, 60-nm

SiGe layer, and 60-nm Si layer).

From the capacitance–voltage characteristics of the samples with Ge/Si(001) heterostructures grown on the KDB12 Si substrates, the surface densities of holes in them were earlier estimated as $3.4 \times 10^{11}$, $7 \times 10^{11}$, and $1.7 \times 10^{11}$ cm$^{-2}$ for $h_{Ge}$ = 6, 10, and 14 Å, respectively. These values almost coincide with the densities of the Ge clusters in arrays (in each layer of quantum dots) [18]. Note also that Raman spectroscopy of these samples indicates that after the deposition of silicon on Ge layers with $h_{Ge}$ from 6 to 10 Å, quantum dots contain (34 ± 2)% Si (estimate over five layers of quantum dots on average) and diffusion of Si into Ge clusters occurs only when a silicon layer was grown over a germanium layer ($T_{gr}$ = 530°C) and does not occur when the germanium layer was grown on the silicon layer ($T_{gr}$ = 350°C) [19]. It is reasonable to assume that the resulting chemical compositions of Ge clusters in different layers of the five-layer structure are different; the Si content is highest in the first (deepest) layer of clusters and is lowest in the fifth layer. This means that potential wells associated with quantum dots are shallower in a deeper located layer of the clusters and the main contribution to the useful signal in the capacitance–voltage characteristics most likely comes from the fifth (closest to the surface) layer of Ge clusters.

The terahertz and sub-terahertz characteristics of the prepared structures were measured at room temperature with a quasioptical spectrometer using backward wave oscillators (BWO) as monochromatic radiation sources continuously tunable in the frequency [20]. A BWO-spectrometer spectrometer is an efficient instrument for investigating terahertz and sub-terahertz electrodynamic properties of various objects from insulators and semiconductors to systems with correlated electrons [21]. In the case under consideration, we measured the spectrum of the transmission coefficient of the silicon substrate with the deposited multilayer heterostructure for THz range, as well as of the part of the substrate from which this heterostructure was removed by etching. All measurements were performed under the conditions of the normal incidence of terahertz radiation on a planar sample, i.e., the magnetic and electric field vectors of radiation lie in the plane of the heterostructure. The typical spectra of the transmission coefficient are shown in Fig. 3. They contain oscillations that are due to the interference of radiation inside the plane–parallel Si substrate, which is an analog of a Fabry–Perot interferometer [22]. The period of the oscillations $\Delta\nu = c/2nd$, where $c$ is the speed of light in vacuum, depends on the thickness $d$ and refractive index $n$ of the substrate and the transmission coefficient in maxima $T_{max}$ is deter mined by the absorption both in the substrate and in the deposited multilayer structure. As is seen in Fig. 3, interference oscillations in the spectrum of the trans mission coefficient of the substrate with the structure are slightly shifted towards low frequencies because of a small (about 1–2 µm) inhomogeneity in the thickness of the substrate. Furthermore, the plotted spectra clearly demonstrate that the $T_{max}$ values of the substrate with the structure are much smaller than the $T_{max}$ values of the substrate without any structure. The measurements of the spectra of the transmission coefficient in various parts of the substrate without Ge/Si layers indicate that the observed differences cannot be attributed to the inhomogeneity of the dielectric characteristics of the substrate (inhomogeneity in the boron concentration). The measurements with the reference sample demonstrate that differences in the transmission coefficients also cannot be explained by any uncontrollable factors of the technological process of the formation of heterostructures; the absorption of radiation in the reference sample is absent. Therefore, the difference between the spectra of the transmission coefficients of the samples with and without the Ge/Si structure is caused by the presence of the layers with Ge quantum dots in this structure. It is worth noting that the effect of additional absorption is present in the case of the presence of Ge quantum dots in these layers, because such absorption is not observed in the measurements with SiGe/Si hetero structures.

Note that additional absorption of terahertz radiation is also not detected in the sample with the Ge/Si heterostructure grown on the high-resistance KDB1000 Si substrate. This implies that the main source of charge carriers that absorb radiation in the studied structures with Ge quantum dots is the initial KDB12 silicon substrate rather than impurities and/or defects in the heterostructure itself (recall that the grown structures were undoped).

The main experimental result of this work is that the absolute absorption coefficient in the heterostructure with Ge quantum dots is much larger than the absorption coefficient of the structure containing almost the same amount of germanium that does not form the array of quantum dots. To describe absorption, we take two characteristics: dynamic conductivity σ and absorption coefficient $α = 4πk/λ$, where $k$ is the attenuation coefficient or the imaginary part of the complex refractive index $n^* = n + ik$ and $λ$ is the wave length of radiation [23]. These quantities σ and α were determined by the procedure of processing spectra measured by a backward wave tube spectrometer [20] using the Fresnel expressions for the transmission coefficient of layered systems [22]. In our case, the system contains 13 layers: the substrate, buffer and cap layers, and five double layers consisting of Ge quantum dots and an Si barrier layer. The dielectric properties of the substrate, buffer, cap, and intermediate layers were preliminarily determined in individual experiments. The solid line in Fig. 3 is an example of the processing of the transmission spectrum measured for the sample with $h_{Ge} = 10$ Å.

Figure 4 shows the final results in the form of the dependence of the conductivity and absorption coefficient of the layers of Ge quantum dots on $h_{Ge}$. In view of large errors (30%) in the determination of the absolute values of σ and α, we cannot reveal possible dependence of these quantities on $h_{Ge}$ for $h_{Ge} < 14$ Å or on the frequency (in the range of 0.03–1.2 THz). Nevertheless, the data presented in Fig. 4 allow two conclusions. First, the dynamic conductivity $σ_{QD} ≈ 90$ $Ω^{-1}cm^{-1}$ and absorption coefficient $α_{QD} ≈ 2900$ $cm^{-1}$ of the layers of quantum dots recalculated on one layer of Ge quantum dots are much larger than the respective values $σ_{Ge} ≈ 0.01$ $Ω^{-1}cm^{-1}$ and $α_{Ge} ≈ 30$ $cm^{-1}$ for the bulk single crystal germanium. Second, a strong decrease in σ and α is observed in the sample with $h_{Ge} = 18$ Å; it is reasonable to attribute this decrease to the weakening of the effect of quantum dots on the dielectric properties of the sample, as the amount of deposited germanium increases and quantum dots coalesce to the continuous nanocrystalline layer (see Fig. 1h). Note that the relative changes in $T_{max}$ introduced by heterostructures (see Fig. 3) are small (in agreement with the known infrared and microwave experiments [24–27]) in view of the smallness of the thicknesses of the layers with Ge quantum dots, calculated changes in their dielectric properties as compared to the properties of bulk germanium are very large, as is seen. It is also worth noting that no absorption of terahertz radiation was observed in the samples with $h_{Ge} = 4.4$ and 6 Å. As in the case of the sample with $h_{Ge} = 18$ Å, this can be explained either by the absence of quantum dots in the Ge layer ($h_{Ge} = 4.4$ Å) or by their small sizes, a large fraction of the free wetting layer, and relatively large distances between the clusters as compared to their sizes ($h_{Ge} = 6$ Å), i.e., by the absence or smallness of the effect of quantum dots on the dielectric properties of the heterostructure.

Both aforementioned conclusions stay valid when the transmission spectra are processed, considering the nanolayers of the heterostructure as having the complex surface conductivity $Y = Y' + iY'' $ or complex surface impedance $R = R' + iR'' = 1/Y$ without the assignment of certain thicknesses to these nanolayers and the specification of their structure. The corresponding model of the complex conducting surface is efficient when investigating the properties of thin inhomogeneous layers [28]. For the case of the homogeneous conducting film whose thickness $d$ is much smaller than the thickness of the skin layer, $Y'' = 0$ and $Y' = 1/R' = 1/σd$ (σ is the conductivity of the film). The application of this model gives the following surface conductivities of the layers of quantum dots and initial germanium: $Y'(QD) ≈ 3.7 × 10^{-3} Ω^{-1} cm^{-1}$, $Y'(Ge) ≈ 4.1 × 10^{-7} Ω^{-1}cm^{-1}$.

The most detailed investigations of microscopic mechanisms of the absorption of infrared and terahertz radiations were performed with GaAs-based nanostructures. The experiments concerning infrared spectroscopy and photoinduced absorption in magnetic fields and without magnetic field exhibit resonance caused by the transitions between quantized energy levels inside quantum dots and quantum wells, as well as between the split levels and the continuum of the valence or conduction band [6, 29–32]. Carriers localized inside quantum dots can form bound states with the carriers in the continuum (excitons) or with LO phonons (polarons), which can in turn interact with each other and form collective complexes [4–6, 25, 30–32]. Plasma excitations generated by electro magnetic radiation in the assembly of conducting clusters or quantum dots also have energies of about 10 meV [33–36].

The listed effects are mainly observed at sufficiently low temperatures. At the same time, there is experimental evidence that these effects influence the properties of the studied objects at higher temperatures, up to room temperature. In particular, the intraband–interlevel absorption of infrared radiation with a frequency of 500–800 cm$^{-1}$ was observed in [31] at 77 and 300 K on arrays of InAs/InAlAs quantum dots grown on the InP(001) substrate. Simulating the kinetics of polariton states in GaN-based microcavities (quantum wells), Malpuech et al. [37] (see also references in that work) showed the possibility of a laser effect on such structures at 300 K. More recently, the emission of coherent radiation was experimentally observed in GaAs/GaAlAs column microcavities at 10 K [38] and in structures with microcavities on GaN quantum wells at 300 K [39].

The experimental data reported above do not allow the certain identification of the mechanism of the absorption of terahertz radiation in Ge/Si hetero structure with Ge quantum dots observed in this work at room temperature. Among the aforementioned mechanisms, the mechanisms of absorption on polaritons and plasma excitations seem to be least affected by thermal fluctuations. The first mechanism is associated with the appearance of the photon–exciton bound state, i.e., with the interaction of the exciton with the photon eigenmode of the microcavity, i.e., quantum dot or quantum well. In contrast to "usual" polaritons in three-dimensional crystals, where the energy of polaritons is proportional to the wave vector, this energy in structures with lower dimensions is finite at zero wave vector and corresponds to terahertz frequencies [37]. The plasmon modes appear because of the localization of carriers in the potential relief created by the geometric boundaries of the elements of nanostructures, surface states, and donor/acceptor ions. The displacement of carriers from the equilibrium position is accompanied by the appearance of a restoring force whose characteristic value corresponds to the presence of plasmon excitations in the terahertz frequency region [7, 24, 27, 34]. In order to reveal the microscopic mechanism of the observed absorption, we are going to perform further investigations of heterostructures with various geometric and physical parameters, as well as in a wider frequency interval (including the infrared range) and at various temperatures.

To summarize, the spectra of the dynamic conductivity (radiation absorption coefficient) in Ge/Si(001) heterostructures with quantum dots have been measured for the first time in the frequency range of 0.3–1.2 THz at room temperature. It has been found that the effective conductivity (absorption coefficient) of the two-dimensional array of germanium quantum dots is several orders of magnitude larger than the conductivity (absorption coefficient) of bulk unstructured single crystal germanium. The possible microscopic mechanisms of the detected increase in the absorption in the array of quantum dots have been discussed.

This work was supported by the Branch of Physical Sciences, Russian Academy of Sciences, basic research program "Modern Problems of Radiophysics," and by the Ministry of Education and Science of the Russian Federation, state contract no. 14.740.11.0069, project no. 20101.1122084032.

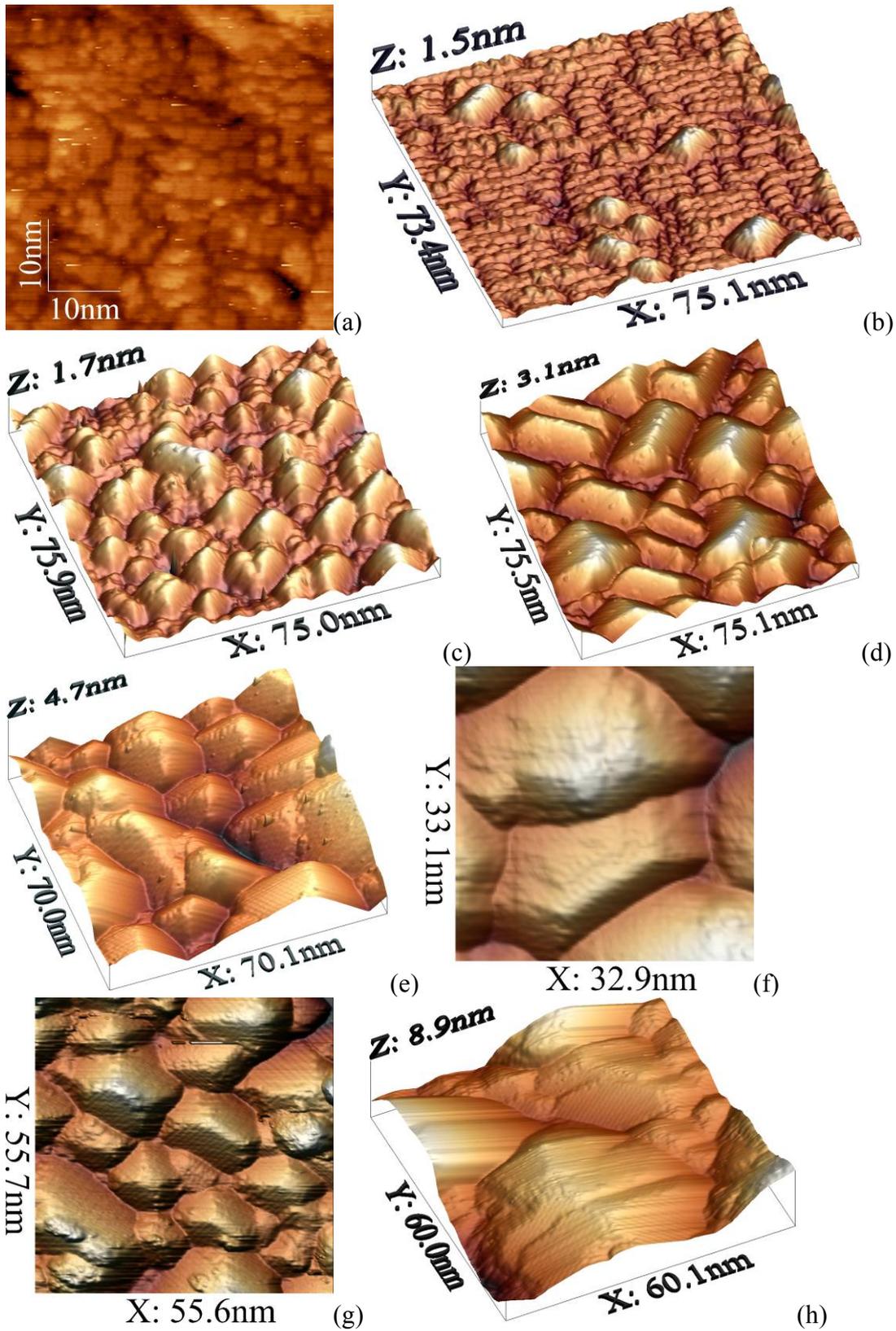

**Fig. 1.** Scanning tunneling microscopy images of the Ge wetting layer before the appearance of an array and arrays of germanium hut clusters at $h_{Ge}$ = (a) 4.4, (b) 6, (c) 8, (d) 10, (f) 14, (g) 15, and (h) 18 Å.

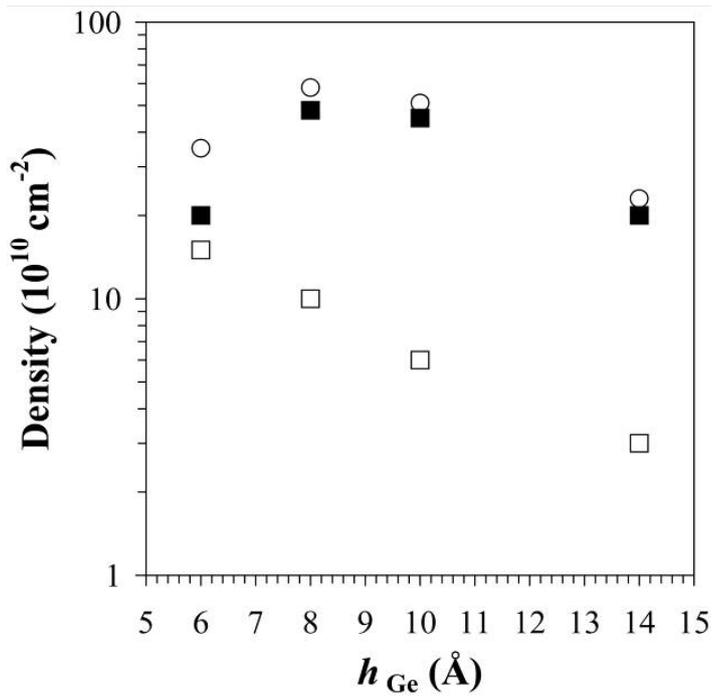

**Fig. 2.** Densities of (□) pyramidal and (■) wedge-shaped Ge hut clusters in arrays versus $h_{Ge}$. The circles present the total density.

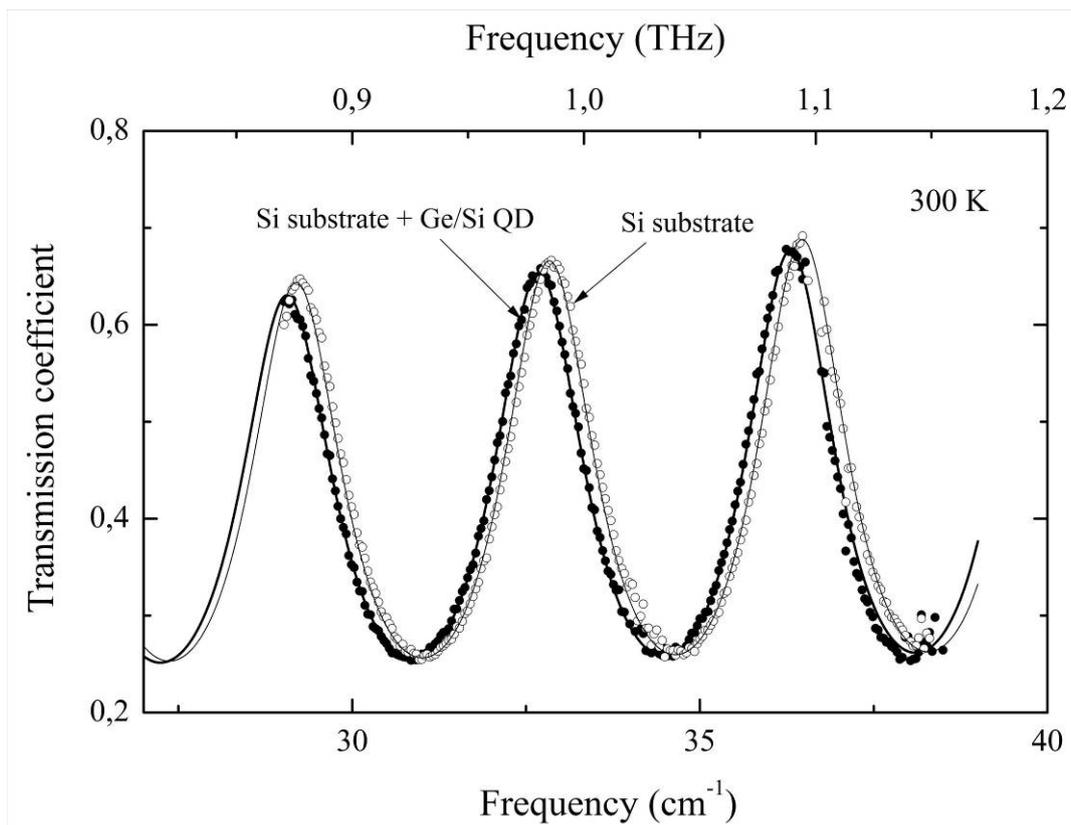

**Fig. 3.** Transmission coefficient spectra of the Ge/Si(001) heterostructure ($h_{Ge}$) measured at room temperature. The points are the experimental data and the line is the least squares approximation with the formula for the transmission coefficient of the 13-layer medium.

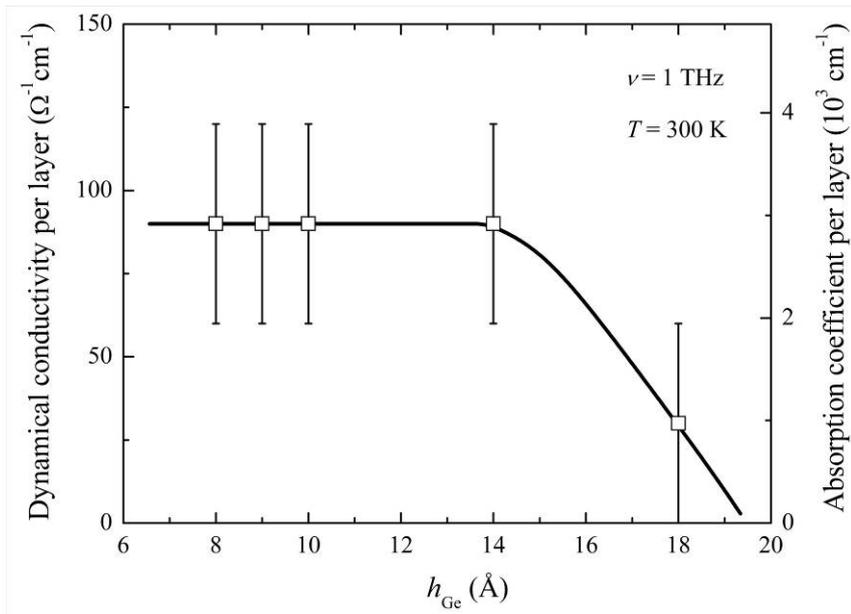

**Fig. 4.** Effective terahertz conductivity σ and effective absorption coefficient α of terahertz radiation in the layer of quantum dots versus $h_{Ge}$.